\documentstyle[12pt]{article}
\begin{document}

\setlength{\textheight}{240mm}
\voffset=-25mm
\baselineskip=20pt plus 2pt
\begin{center}

{\large \bf On the Difference of Energy between the Einstein and 
M{\o}ller Prescription }\\
\vspace{5mm}
\vspace{5mm}
I-Ching Yang$^{\dagger}$ \footnote{E-mail:icyang@cc.ntttc.edu.tw}
and Irina Radinschi$^{\ddagger}$ \footnote{E-mail:iradinsc@phys.tuiasi.ro}

$^{\dagger}$Department of Natural Science Education and \\
Advanced Science and Technology Research Center, \\
National Taitung Teachers College, Taitung, Taiwan 950, Republic of China \\ 
and \\
$^{\ddagger}$Department of Physics, ``Gh. Asachi" Technical University, \\
Iasi, 6600, Romania

\end{center}
\vspace{5mm}

\begin{center}
{\bf ABSTRACT}
\end{center}
In some black hole solutions, these do not exist the same energy-momentum complexes
associated with using definition of Einstein and M{\o}ller in given coordinates.  
Here, we consider the difference of energy between the Einstein and M{\o}ller 
prescription, and compare it with the energy density of those black hole solutions.  
We found out a special relation between the difference of energy between the 
Einstein and M{\o}ller prescription and the energy density for considered
black hole solutions.

\vspace{2mm}

\newpage
In the theory of general relativity, many physicists, like Einstein~\cite{1}, Landau 
and Lifshitz~\cite{2}, Tolman~\cite{3}, Papapetrou~\cite{4}, M{\o}ller~\cite{5}, and 
Weinberg~\cite{6}, had given different definitions for the energy-momentum complex.
Specifically, the M{\o}ller energy-momentum complex allows to compute the energy in 
any spatial coordinate system.  Some results recently obtained~\cite{7,8,9,10} sustain 
that the M{\o}ller energy-momentum complex is a good tool for obtaining the energy 
distribution in a given space-time. Also, in his recent paper, Lessner~\cite{11} 
gave his opinion that the M{\o}ller definition is a powerful concept of energy and 
momentum in general relativity.  In his paper Virbhadra~\cite{12} point out that 
several energy-momentum complexes (ELLPW) give the same result for a general 
non-static spherically symmetric space-time of the Kerr-Schild class.

In particular, whatever coordinates do not exist the same energy complexes
associated with using definitions of Einstein and M{\o}ller in some space-time 
solutions~\cite{7,13}. According to the definition, the Einstein energy complex 
is~\cite{1}
\begin{equation}
E_{\rm Ein} = \frac{1}{16\pi} \int \frac{\partial H_0^{\;\,0l}}
{\partial x^{l}} d^3 x ,
\end{equation}
where 
\begin{equation}
H_0^{\;\,0l} = \frac{g_{00}}{\sqrt{-g}} \frac{\partial}{\partial x^m} 
[(-g) g^{00} g^{lm} ] ,
\end{equation}
and the M{\o}ller energy complex is~\cite{5}
\begin{equation}
E_{\rm M{\o}l} = \frac{1}{8\pi} \int \frac{\partial \chi_0^{\;\,0l}}
{\partial x^{l}} d^3 x ,
\end{equation}
where 
\begin{equation}
\chi_0^{\;\,0l} = \sqrt{-g} g^{0\beta} g^{l\alpha} \left( \frac{
\partial g_{0\alpha}}{\partial x^{\beta}} - \frac{\partial g_{0
\beta}}{\partial x^{\alpha}} \right) .
\end{equation}
Where the Latin indices take values from 1 to 3, and the Greek indices run from 
0 to 3.  Let us look into the difference of energy between the Einstein and 
M{\o}ller prescription, which be defined as 
\begin{equation}
\Delta E = E_{\rm Ein} - E_{\rm M{\o}l}
\end{equation}
In this article, we would discuss the problem within the difference
between Einstein and M{\o}ller energy-momentum complexes.

In the first case, we think of two solutions of Einstein vacuum 
field equation: \\
(i) Schwarzschild space-time \\
The metric form of Schwarzschild space-time is 
\begin{equation}
ds^2 = fdt^2 - f^{-1} dr^2 - r^2 d\theta^2 - r^2 \sin^2 \theta d\varphi^2, 
\end{equation}
where $ f=1 - 2M/r $.
It is a well-known results that the energy complexes of Einstein and
M{\o}ller of Schwarzschild space-time are
\begin{eqnarray}
E_{\rm Ein} & = & M ,\\
E_{\rm M{\o}l} & = & M  ,
\end{eqnarray}
and the difference is
\begin{equation}
\Delta E =0 .
\end{equation}
(ii) Kerr solution \\
The metric form of Kerr solution is considered as 
\begin{equation}
ds^2 = \alpha dt^2 - \beta dr^2 - \gamma d\theta^2 - \delta d\phi^2 - 2 
\sigma dt d\varphi, 
\end{equation}
where $\alpha = 1 - 2Mr / \Sigma$ , $\beta = \Sigma / \Delta$ , $\gamma = 
\Sigma$ , $\delta = r^2 + a^2 + 2Ma^2 r \sin^2 \theta / \Sigma$ and
$\sigma = 2Mar \sin^2 \theta / \Sigma$ . Here $\Sigma \equiv r^2 + a^2
\cos^2 \theta$ and $\Delta \equiv r^2 - 2Mr + a^2$. 
To use the results in the Virbhadra articles~\cite{14} and to set these 
$Q=0$, we could obtain the enrgy-momentum complexes of Einstein and
M{\o}ller of Kerr space-time are
\begin{eqnarray}
E_{\rm Ein} & = & M , \\
E_{\rm M{\o}l} & = & M ,
\end{eqnarray}
and the difference is
\begin{equation}
\Delta E = 0 .
\end{equation}
For Einstein's vacuum field equation, the energy density is
\begin{equation}
T_0^{\;\,0} = 0 .
\end{equation}
We would find that $\Delta E$ equal to the value of $T_0^{\;\,0}$.

Next, we consider two case of the coupled system of the Einstein field 
and electromagnetic field: \\
(iii) Reissner-Nordstr\"{o}m space-time \\
The metric form of Reissner-Nordstr\"{o}m space-time is
\begin{equation}
ds^2 = fdt^2 - f^{-1} dr^2 - r^2 d\theta^2 - r^2 \sin^2 \theta d\varphi^2, 
\end{equation}
where $ f=1 - 2M/r + Q^2/r^2$.
Previously, the energy-momentum complexes of Einstein and M{\o}ller of
Reissner-Nordstr\"{o}m space-time had been calculated with
\begin{eqnarray}
E_{\rm Ein} & = & M - \frac{Q^2}{2r} ,\\
E_{\rm M{\o}l} & = & M - \frac{Q^2}{r} ,
\end{eqnarray}
and the difference is 
\begin{equation}
\Delta E = \frac{Q^2}{2r} .
\end{equation}
Notice that the energy density in the Einstein-Maxwell field equation is
\begin{equation}
T_0^{\;\,0} = \frac{Q^2}{r^4}. 
\end{equation}
(iv) charged regular black hole \\
The metric form of charged regular black hole is~\cite{15} 
\begin{equation}
ds^2 = fdt^2 - f^{-1} dr^2 - r^2 d\theta^2 - r^2 \sin^2 \theta d\varphi^2, 
\end{equation}
where $ f=1- \frac{2M}{r}(1-{\rm tanh}(\frac{Q^2}{2Mr})) $.
Using the results of Radinschi's articles~\cite{10}, the energy-momentum 
complexes of Einstein and M{\o}ller of charged regular black hole are
\begin{eqnarray}
E_{\rm Ein} & = & M \left[ 1- \tanh(\frac{Q^2}{2Mr}) \right] ,\\
E_{\rm M{\o}l} & = & M \left[ 1- \tanh(\frac{Q^2}{2Mr}) \right] 
 - \frac{Q^2}{2r} \left[ 1- \tanh^2(\frac{Q^2}{2Mr}) \right] ,
\end{eqnarray}
and the difference is
\begin{equation}
\Delta E = \frac{Q^2}{2r} \left[ 1- \tanh^2(\frac{Q^2}{2Mr}) \right] .
\end{equation}
However, the energy density of the coupled system of the Einstein field
and nonlinear electrodynamics field is
\begin{equation}
T_0^{\;\,0} = \frac{Q^2}{r^4} \left[ 1- \tanh^2(\frac{Q^2}{2Mr}) 
\right] .
\end{equation}
Here the relation between 
$\Delta E$ and the energy density is written as
\begin{equation}
\Delta E = T_0^{\;\,0} \times (\frac{r^3}{2}) .
\end{equation}

Then, in the last case, we use a special black hole solution without 
singularity:\\
(v) static spherically symmetric nonsingular black hole \\
The metric form of the static spherically symmetric nonsingular black 
hole solution~\cite{16} is
\begin{equation}
ds^2 = fdt^2 - f^{-1} dr^2 - r^2 d\theta^2 - r^2 \sin^2 \theta d\phi^2, 
\end{equation}
where $ f=1 - R_g/r $ and $R_g = r_g (1 - \exp(r^3/r_*^3))$ , $r_*^3 = 
r_g r_0^2$ ,$r_g = 2M$ , $r_0^2  = \frac{3}{8\pi \varepsilon_0}$ .
According to the results of our articles~\cite{17}, the energy-momentum 
complexes of Einstein and M{\o}ller of the static spherically symmetric 
nonsingular black hole are
\begin{eqnarray}
E_{\rm Ein} & = & M - M \exp ( -\frac{r^3}{r_*^3} ) ,\\
E_{\rm M{\o}l} & = & M - M \exp ( -\frac{r^3}{r_*^3} ) -
\frac{3r^3}{r_0^2} \exp ( -\frac{r^3}{r_*^3} ) ,
\end{eqnarray}
and the difference is 
\begin{equation}
\Delta E = \frac{3r^3}{r_0^2} \exp ( -\frac{r^3}{r_*^3} ) .
\end{equation}
Notice that the energy density of the static spherically symmetric 
nonsingular black hole be assumed as 
\begin{equation}
T_0^{\;\,0} = \frac{3}{r_0^2} \exp ( -\frac{r^3}{r_*^3} ) .
\end{equation}
The relation between $\Delta E$ and the energy density is written as
\begin{equation}
\Delta E = T_0^{\;\,0} \times r^3 .
\end{equation}

Although, we could summarize that the general relation between $\Delta E$ 
and the energy density $T_0^{\;\,0}$ be written as
\begin{equation}
\Delta E = T_0^{\;\,0} \times (kr^3) ,
\end{equation}
with $k=1/2$ and $k=1$.  But, it is still an open question why the special 
relation has between $\Delta E$ and the energy density $T_0^{\;\,0}$. 
Further study is needed to understand the difference between the Einstein 
and M{\o}ller energy complexes of more 
varied black hole solutions. 

\begin{center}
{\bf Acknowledgements}
\end{center}
I.-C. Yang thanks the National Science Council of the 
Republic of China for financial support under the contract number 
NSC 90-2112-M-143-003.

\end{document}